\documentstyle[version2,aps,epsf,titlepage]{revtex}
\draft
\begin{document}
\bibliographystyle{prsty-all_names}
\twocolumn[{
\begin{title}
Kondo Effect in a Metal with Correlated Conduction Electrons 
\end{title}
\author{S. Tornow$^1$, V. Zevin$^2$, and G. Zwicknagl$^1$}
\begin{instit}
$^1$Max-Planck-Institut f\"ur Physik Komplexer Systeme,
Au{\ss}enstelle Stuttgart,\\ Postfach 80 06 65, 70506 Stuttgart, Germany \\
$^2$The Racah Institute of Physics, The Hebrew
University of Jerusalem, 91904 Jerusalem, Israel
\end{instit}
\begin{abstract}
The large-degeneracy expansion for dilute magnetic alloys is extended 
to account for conduction electrons interactions. Particular attention
is paid to the renormalization of the hybridization vertex which
affects the low-energy excitations. As a first example, we calculate 
the enhanced characteristic energy $k_BT_0$ in the limit of weakly 
correlated conduction electrons. The metallic regime with strongly 
correlated electrons is discussed.
\end{abstract}
}]
\narrowtext
Correlations among the conduction electrons may strongly affect the
ground state and the low-energy excitations of metals with magnetic
ions. A typical example are doped cuprates with rare earth ions
Nd$_{2-x}$Ce$_x$CuO$_4$
where a novel type of heavy fermion-like behavior was recently
discovered \cite{Bruggeretal}. Although the thermodynamic properties in the
normal state closely parallel the behavior of standard heavy fermion
compounds \cite{GreweSteglich,Ottreview} the observed characteristic 
energy of the low-energy excitations cannot be related to a 
Kondo temperature estimated
from the known coupling between the spins of the rare earth ions and
the conduction electrons \cite{FZZ,TZZ}. 

In this paper, we study the influence of conduction electron
correlations on the low-energy excitations
of dilute magnetic alloys. The latter are
characterized by an energy scale $k_B T_0$ much smaller than
the characteristic energies of the conduction electrons such as the
bandwidth D or the local Coulomb repulsion $U$. We thereby tacitly assume that
the conduction electron correlations do not introduce low-energy
anomalies i.e. that the conduction electron properties are smooth on
the scale set by $k_B T_0$. This separation of energy scales forms the
basis of our theory. It allows us to reduce the general problem of a
magnetic impurity in a metal with strongly correlated electrons to a
form which can be solved in close analogy to the well-studied case of
uncorrelated conduction electrons.  
Of particular interest is the characteristic energy $k_BT_0$
which in the case of uncorrelated conduction electrons depends 
exponentially upon the inverse of the coupling 
between localized and conduction electrons 
\cite{TsvelickWiegmann,FKZreview,AdvPhysReview,HewsonBook}. 
This exponential many-body 
scale is a direct consequence of the existence of Fermi liquid low frequency 
excitations in a normal metal. Theoretical studies of an impurity 
spin embedded in a one-dimensional Luttinger liquid 
\cite{LeeToner,FurusakiNagaosa,Avi}, on the other hand, predict a
power law for the variation of the characteristic temperature with the coupling
in the limit of large on-site Coulomb interaction U.

The modifications introduced by the conduction electron interactions
into the low-energy excitations arise from the subtle interplay of
three different types of influences. First, the density of conduction
states at the Fermi level is changed. Second, the virtual transitions
between f- and conduction states are reduced. Third, the effective
spin coupling between the conduction and f-electrons is enhanced by
the increased number of uncompensated spins in the correlated
conduction electron system. Considering these facts, it is not
surprising that model studies accounting only for selected aspects
arrive at rather controversial conclusions concerning the Kondo effect
in metals with correlated electrons 
\cite{KhaliullinFulde,ItaiFazekas,Igarashi,Schork}. 

For a microscopic description, we consider an Anderson impurity 
coupled to interacting electrons. The latter are described by a one-band 
Hubbard model. The resulting Hamiltonian reads
\begin{eqnarray}
H&=&\sum_{{\underline k} \sigma} \epsilon_{\underline k} 
c_{{\underline k} \sigma}^{\dagger} 
c_{{\underline k} \sigma} + {U\over2}
\sum_{{\underline k}, {\underline {k^{\prime}}}, 
{\underline q }\sigma \neq \sigma^{\prime}} 
c_{{{\underline k}+{\underline q}} \sigma}
^\dagger c_{{{\underline k^{\prime}}-{\underline q}} \sigma^{\prime}}^\dagger 
c_{{\underline k^{\prime}} \sigma^{\prime}} 
c_{{\underline k} \sigma} \nonumber\\
&+&\sum_{m} \epsilon_f n_{fm} +
{U_f\over2}\sum_{m \neq m{\prime}}n_{fm} n_{f m^{\prime}}\nonumber\\
&+&\sum_{{\underline k}, m ,\sigma} 
\left(V_{{\underline k} m \sigma} f_{m}^{\dagger}
      c_{{\underline k} \sigma}+h.c.\right)     
\label{eq:H}
\end{eqnarray}
The operators 
$c_{{\underline k} \sigma}^{\dagger}(c_{{\underline k} \sigma})$
create (annihilate) conduction
electrons with momentum ${\underline k}$, band energy 
$\epsilon_{\underline k}$ and spin $\sigma$. The local Coulomb
repulsion between two conduction electrons at the same site is $U$. The
$f_{m}^{\dagger}(f_{m})$ are the creation (annihilation) operators for
$f$-electrons on the impurity site. They are characterized by the
total angular momentum $J$ and a quantum number $m$
which denotes the different states $m=1,\dots,N_f$ within the
$N_f$-fold degenerate ground state multiplet with orbital energy 
$\epsilon_f$. All energies are measured relative to the Fermi level.
The Coulomb repulsion $U_f$ between two
$f$-electrons at the impurity site is assumed to be much larger than
the other energy scales and therefore we may let 
$U_f \rightarrow \infty$.
We consider here an orbitally non-degenerate Anderson impurity 
putting in Eq. \ (\ref {eq:H}) 
$m=\sigma$ and the hybridization coupling $V_{{\underline k} m \sigma} 
= V$.

The excitation spectra are calculated from the Green's functions of the empty state
$|0>$ (i.e. the $f^0$ or $f^{14}$ configuration) and 
the occupied  $f$ states $|\sigma>$, respectively,
denoted by $G_0(z)$ and $G_\sigma(z)$
\begin{equation}
G_\alpha(z) \;=\;  {1 \over {z- \epsilon_\alpha - \Sigma_\alpha(z)}}
\end{equation}
where $\epsilon_\alpha = 0,\epsilon_f$ for $\alpha=0, \sigma$ respectively .
The interactions among the conduction electrons affect the 
self-energies $\Sigma_0$ and $\Sigma_\sigma$ which are coupled. They
are determined by the self-consistent large degeneracy expansion in
close analogy to the Non Crossing Approximation \cite{BickersRMP} for 
non-interacting conduction electrons.

The derivation of the generalized NCA equations 
procedes in two steps starting from a
conventional perturbation expansion for the self-energies $\Sigma_0$
and $\Sigma_\sigma$ in terms of the bare conduction electron propagators,
the bare Green's functions for the relevant f-configurations, the Coulomb
interaction $U$ and the hybridization $V$. By reordering and partial
summation the series is converted into an expansion in terms of
bare Green's functions for the f-configurations, Coulomb-renormalized 
propagators for the conduction electrons and effective 
hybridization vertices which
account for the correlations among the conduction electrons. 
The resulting diagrams are classified with respect to their order in
the inverse degeneracy in close analogy to the case
$U=0$. Self-consistent summation of the leading terms yields the
self-energies displayed in
Figures \ref{fig:bubble_self}\,a,b for the empty f-state 
and occupied f-states, respectively,while Figure
\ref{fig:bubble_self}\,c 
illustrates the general structure of the effective hybridization 
vertex for the case of two-particle correlations. 
The correlation-induced vertex corrections generally contain an 
n-electron Green's function where $2n-1$ external lines are 
connected by $n$ Green's functions
for empty and occupied f-configurations, respectively, via $2n-1$ bare
hybridization vertices. On the relevant low-energy scale the variation
with energy of the effective vertices and of the selfenergies is
determined by the structure of the diagrams and the Green's functions $G_0$
and $G_m$. This fact follows directly from the separation of energy
scales. For the Hubbard model, the dominant contribution originates
from two-particle correlations. We shall therefore restrict ourselves
to the self-consistency equations displayed in Figure {\ref{fig:bubble_self}}. 

Neglecting the vertex corrections from the Coulomb interaction yields
self-energies
\begin{eqnarray}
\Sigma_0^{(0)}\left(\omega \right) = \frac{V^2}{N}\sum_{{\underline k}
\sigma}\!\! \int d\xi
n_{F}(\xi) A_{\sigma} ({\underline k},\xi) G_{\sigma}\left(\omega + \xi\right)
\nonumber \\
\Sigma_{\sigma}^{(0)}\left(\omega \right) = \frac{V^2}{N}
\sum_{{\underline k}}\!\! \int d\xi
n_{F}(-\xi) A_{\sigma} ({\underline k},\xi) G_{0}\left(\omega - \xi\right) 
\label{eq:bubble_self}
\end{eqnarray}
in close analogy to the case of non-interacting electrons \cite {ColemanNCA}.
In Eq. ({\ref{eq:bubble_self}}),  $n_{F}(\xi)$ is the Fermi function while  
$A_{\sigma}({\underline k},\xi)$ denotes the spectral function of interacting 
electrons which for $U=0$, reduces to
$\delta \left(\xi - \epsilon_{\underline k}\right)$. 
For non-interacting conduction electrons, the self-consistent solution
has three characteristic features: The occupied f-spectrum
shifts to peak at a value $E_f$ the dominant contribution to the
level shift coming from the continuum of charge fluctuations. The 
resonance in the occupied f-spectrum acquires a small width. Finally, 
the empty state spectral
function exhibits a pronounced structure at $\omega_0=E_f-k_BT_0$
which develops with decreasing temperature and which sets the scale
for the low-temperature behavior. This feature is the direct manifestation of
the Kondo effect reflecting the admixture of $f^0$-contributions to
the ground state and the low-energy excitations.
 
It is obvious \cite{promises} that also for interacting conduction
electrons the dominant effect of hybridization on the $4f^1$
configurational spectrum is a shift $E_f(U)-\epsilon_f=Re \Sigma_\sigma$
of the resonance energy which, however, is renormalized by
the Coulomb repulsion U and its influence
on the charge fluctuations. 
The interaction-induced shift, however, is rather small
\cite{promises} and will be neglected in the subsequent discussion.
The central quantity to be studied here is
the empty state self energy and, in particular, its variation with
energy close to $E_f$ which can be deduced form rather
general considerations. The smooth variation with energy of 
$\sum_{{\underline k}} A_{\sigma} ({\underline k},\xi)$ implies that the
basic analytic structure of $\Sigma_0^{(0)}(\omega)$ is not altered
as compared to the case of non-interacting conduction electrons, the
characteristic feature being a logarithmic variation in the vicinity
of the f-energy $E_f$. The prefactor, however, is
proportional to the interaction-renormalized density of
states at the Fermi level $\rho(0)$. The low-energy scale, $k_BT_0$, 
i.e. the distance between the pole in
the empty f-state Green's function and the $4f^1$ peak depends 
on the renormalized parameters in the usual exponential way.
The vertex correction displayed in Figure {\ref{fig:bubble_self}\,c}
accounts for the two-particle cocrrelations in the interacting conduction
electron system. The corresponding contribution to the empty state
self energy is denoted by      
$\Sigma_0^{(1)}\left(\omega\right)$. We should like to emphasize that
the latter is proportional to
$V^4N_f \left( N_f - 1\right)$ and hence
of the same order in the inverse local degeneracy 
${\left( 1/N_f \right)}^0$ as the leading term 
$\Sigma_0^{(0)} \left(\omega\right)$. The same classification applies
to the $4f^1$ configurational self energies 
$\Sigma_\sigma^{(0)}$ and $\Sigma_\sigma^{(1)}$ which are
both of the order ${\left( 1/N_f \right)}^1$. 

In this paper we shall evaluate and discuss the contribution to
lowest order in the effective interaction. The conclusions we shall 
arrive at can easily be extended to intermediate values of U by
inserting a more sophisticated approximation to the 
appropriate two-particle t-matrix. However, the
latter cannot be simply expressed in terms of Landau parameters since
the microscopic expression involves conduction electron energies far
from the Fermi surface. To study the analytic behavior in the energy
range of interest, we insert the unperturbed
conduction electron propagator and obtain
\begin{eqnarray}
& &\Sigma_0^{(1)} (\omega) =-\frac{2 U V^4}{N^3}  
\!\sum_{{\underline k},{\underline k}^{\prime},{\underline q}}
\frac{g_{{\underline k},{\underline k}^{\prime},{\underline q}}-g_{
{\underline k}, {\underline k^{\prime}} ,{\underline q} }}
{\epsilon_{
\underline k}-\epsilon_{\underline k^{\prime}} 
+ \epsilon_{\underline k^{\prime}
+ {\underline q}} - \epsilon_{{\underline k} + {\underline q}}}  
\label{eq:U_self}
\\ \nonumber
& &g_{{\underline k},{\underline k^{\prime}}, 
{\underline q}}
= G_0(\omega + 
\epsilon_{\underline k} 
-\epsilon_{{\underline k} + {\underline q}}) G_\sigma(\omega
  + \epsilon_{\underline k}) n_{F}(\epsilon_{\underline k})n_F(-\epsilon_{
{\underline k} + {\underline q}}) \nonumber \\ \nonumber
& &[n_{F}(\epsilon_{{\underline k}^{\prime}})G_\sigma(\omega\!+\!\epsilon_{
\underline k^{\prime}})\!-\!n_{F}(\epsilon_{{\underline k}^{\prime}+{\underline q}
}) G_\sigma(\omega+\epsilon_{
\underline k}\!-\!\epsilon_{{\underline k}+{\underline 
q}}\!+\!\epsilon_{{\underline k^{\prime}}+{\underline q}})]      
\end{eqnarray}
The occupied state self-energy $\Sigma_{\sigma}^{(1)}$ is expressed 
analogically \cite{analogy}.
To make things simpler we calculate $\Sigma_0^{(1)}\left(\omega\right)$ 
for $T = 0$ in the local approximation, hence neglecting the
momentum conservation in Eq. (\ref{eq:U_self}), and obtain     
\begin{eqnarray}
\Sigma_0^{(1)} \left(\omega \right)& = & 
-\frac{1}{2 \pi} U \rho(0) \frac{\Gamma}{D}
I_0^{(1)}(\omega)\Sigma_0^{(0)}(\omega) \nonumber \\ 
 & &-\frac{1}{2 \pi^2} U 
({\frac{\Gamma}{D}})^2  I_0^{(2)}(\omega)
\label{eq:Sigma1_T=0}
\end{eqnarray}
Here 
$\Sigma_0^{(0)} \left( \omega \right)$ = 
$\frac{2 \Gamma}{\pi} 
\ln{\left|\frac{\omega-\epsilon_f}{\omega -\epsilon_f-D}\right|}$ 
and
$I_0^{(1)}$ and $I_0^{(2)}$ are dimensionless
functions of $\omega/D$. For the square DOS and half-filling $I_0^{(1)}$ is
given by the following equation
\begin{equation}
I_0^{(1)}(\omega)\;=\;\int_{-D}^{0}\!\!d\xi\!\!\int_{0}^{D}\!\!d\eta
\!\!\int_{-D}^{0}\!\!d\zeta 
\frac{2 \ \ G_0
(\omega\!+\!\xi\!-\!\eta)}{(\omega\!-\!\epsilon_f\!+
\!\xi)(\omega\!-\! \epsilon_f\!+\!\xi\!-\!\eta\!+\!\zeta)}
\label{eq:I_1}
\end{equation}
A similar expression is obtained
for the second integral, $I_0^{(2)}$ \cite{comments}.
Indeed on this first
iteration stage of the NCA equations we already face an integral equation
for the empty f-state Green's function ( see Eq. (\ref{eq:I_1}) ). 
A detailed analysis shows that in the energy range of interest the
integrals vary approximately like 
$I_0^{(1)}(\omega) \sim A_1 \ln |\omega-\epsilon_f| + B_1$ and
$I_0^{(2)}(\omega) \sim -(A_2 \ln |\omega-\epsilon_f| + B_2)^2$
where the values of the constants $A_i>0,B_i>0,i=1,2$ depend on the shape
of the empty state spectral function. Detailled expressions will be
given in a forthcoming paper.
For a first quantitative estimate of the intgrals $I_{1,2}$ 
we use for $G_0(\omega)$ in 
Eq. (\ref{eq:I_1}) the simplest possible form which accounts for 
the charge fluctuations only. This can be justified in the 
Kondo-regime where the spectral weight of the charge fluctuations is
of order unity. In addition, it can be shown that the spin
fluctuations do not introduce new divergences \cite{promises}. 
The full NCA iteration procedure will be discussed 
elsewhere \cite{promises}. The numerical solution for the characteristic 
temperature $k_BT_0=\epsilon_f-\omega_0$ with $\omega_0$ given by 
\begin{equation}
\omega_0-\Sigma_0^{(0)}(\omega_0)-\Sigma_0^{(1)}(\omega_0)\;=\;0
\label{eq:pole}
\end{equation}
is displayed Figure \ref{fig:T_Kgraphs} which 
compares $T_0 \left( U \right)$ as function of $\Gamma$ for 
given U and $\epsilon_f$. The conduction electron repulsion 
leads to a net enhancement of $T_0(U)$ as
compared to the non-interacting case. 

To better understand this result we evaluate the vertex corrections at
the fixed energy $\omega_0=\epsilon_f-T_0(0)$. This "on-shell" approximation
which was adopted in \cite{KhaliullinFulde} leads to the renormalization
\begin{eqnarray}
& &\Gamma \rightarrow \Gamma \left( 1 - \frac{U}{2 \pi} 
\rho\left(0\right)
\frac{\Gamma}{D} I_0^{(1)} \left(\omega_{0}\right) \right) 
\label{eq:renrmlztn} \\ \nonumber
& &\epsilon_f \rightarrow \epsilon_f + \frac{ U}{2 \pi^2}  
({\frac{\Gamma}{D}})^2 I_0^{(2)} \left(\omega_0 \right)
\end{eqnarray} 
Both integrals $I_{1,2} \left(\omega_0 \right)$ are negative,
so the former renormalization leads to an increase of $T_0 \left( U \right)$ 
whereas the latter leads to a decrease ( see also \cite{Schork} ).
Calculations of integrals in
Eq. (\ref{eq:renrmlztn}) show that the $\Gamma$-renormalization wins and 
$T_0 \left( U \right)$ increases with U in agreement with 
\cite{KhaliullinFulde}. 
The "on-shell" results for $T_0(U)$ which are included in Figure
{\ref{fig:T_Kgraphs}} for comparison tend to overestimate the
renormalization.
 
The on-shell approximation
in Eq.(\ref{eq:renrmlztn}) may be represented as $\Gamma_{eff}
 = \Gamma [ 1+ CU\rho\left( 0 \right)]$ where C is 
itself a function of $\Gamma$ and $\epsilon_f$. For $\epsilon_f/D 
= -0.67$ and $0.16 \leq \Gamma/D \leq 0.3$ we obtain 
$0.72 \geq \ C  \geq 0.54$
These values has to 
be compared with the $C \approx 1$ in the Kondo spin model 
of \cite{KhaliullinFulde}.
Analogically we obtain $\epsilon_f = \epsilon_f \left( 1 + C_1 U/|E_f|\right)$
with $0.02 \geq C_1 \geq 0.016$.

For weak electron-electron interaction the increase of $T_0$ may be 
understood as resulting from the reduced probability of finding doubly
occupied and empty lattice sites in the correlated conduction electron
systems. The increased number of uncompensated conduction electron
spins finally leads to the enhancement of the 
effective hybridization coupling. We may expect that for sufficiently 
large U and a half-filled conduction band the virtual transition from 
the f-state to the conduction state will cost too
much energy inhibiting the $T_0$-increase and leading 
eventually to the change  in the trend. 
The analogy between the Kondo spin model and the Anderson impurity model in its
local moment regime is not complete in the case of correlated
conduction electrons ( see also \cite{Li} ).
In the former case $T_0$ will increase monotonously with the U because
of the enhancement of the exchange interaction. In the latter case the process
is two-staged, it involves the formation of a local moment and its interaction
with the conduction electrons.

The present results are based on the separation of energy scales. We
therefore anticipate no qualitative changes when using 
more sophisticated approximations for the vertices 
$\Gamma_{\sigma,\bar{\sigma}}^{\left( U \right)}\left( 1,2;3,4 \right)$, 
(Figure \ref{fig:bubble_self}\ c) appropriate for the strong correlation
regime. For a quantitative treatment of this problem there are two visible
approaches.  One is to introduce 
summation of the infinite subseries of the RPA and the ladder type. In this 
case we introduce in
self-energies $\Sigma_0^{(1)}\left(\omega\right)$ and $
\Sigma_{\sigma}^{(1)}\left(\omega\right)$
dynamical Lindhard functions. The other way is to approximate the vertex
correction by response functions ( dynamic susceptibilities ). This may be done
with the use of the local approximation 
\cite{ZlaticHorvatic,Jarrel,SIAMreview}.
The work in these directions is in progress\cite{promises}. 
In addition, the influence of conduction electron interactions
on the spectral properties of magnetic impurities and their dependence
upon the doping are also interesting topics for future investigations.

In conclusion, the general NCA equations for the interacting
conduction electrons are obtained and it is
shown that due to the renormalisation of the hybridization interaction the
characteristic energy is increased by the weak interactions.

This work was supported by the Bundesministerium f\"ur Bildung, 
Wissenschaft, Forschung und Technologie through F\"orderprogramm 
``Elektronische Korrelationen und Magnetismus'' (S. T.), the Israel Science 
Foundation  administered  by the Israel Academy of Sciences and Humanities
(V. Z.) and
the Deutsche Forschungsgemeinschaft through Sonderforschungsbereich 252 
``Elektronisch hochkorrelierte metallische Materialien'' 
Darmstadt-Frankfurt-Mainz (G. Z.). The 
authors are grateful to Prof. G. Khaliullin 
for useful comments about vertex corrections and acknowledge discussions with
Prof. P. Fulde. 
S.T. thanks the Racah Institute for hospitality.
The hospitality at MPI PKS is acknowledged by V.Z..

\figure{Self-consistent f configuration self-energies and
contributions to the vertex. The solid, dashed and wavy lines
represent the dressed propagators for conduction electrons, occupied
and empty f states. The open circle denotes the bare hybridization V
while open and filled squares are the bare on-site Coulomb repulsion
and the two-particle vertex 
$\Gamma_{\sigma,\bar{\sigma}}^{(U)}(1,2;3,4 )$, respectively. %
(a) Empty state self-energy $\Sigma_0(i\nu_m)$. %
(b) Occupied state self-energy $\Sigma_\sigma(i\omega_n)$. %
(c) Contribution to the effective hybridization vertex.
(d) Lowest order correction. 
\label{fig:bubble_self}}
\figure{$T_{K}(U)$ as function of $\Gamma$ for $U/D=0.5$ and 
$\epsilon_f/D=-0.67$;. The solid refers to the solutions of
Eq.(\ref{eq:pole}) while the dashed line is for the on-shell
approximation, 
Eq. (\ref{eq:renrmlztn}); 
We also include the U=0 result for comparison (dot-dashed line).
\label{fig:T_Kgraphs}}

\begin{thebibliography}{10}
\bibitem{analogy}
$\Sigma_{\sigma}^{(1)}$ may be obtained from the $\Sigma_{0}^{(1)}$ by omitting
the factor of 2, by changing in the function $g$ superscripts, $b 
\leftrightarrow f$, and by inverting signs of all $\epsilon$ in this function.
\bibitem{comments}
Our integral functions $I_0^{(1)}$ and $I_0^{2}$ are very close to analogical
  integrals in ref. \cite{Schork}, $-I_A^{(1)}$ and the sum $-( I_B^{(1)} +
  I_D^{(1)} )$ correspondengly.
\end{thebibliography}
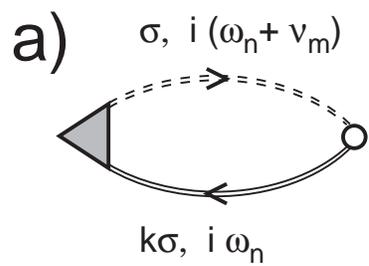
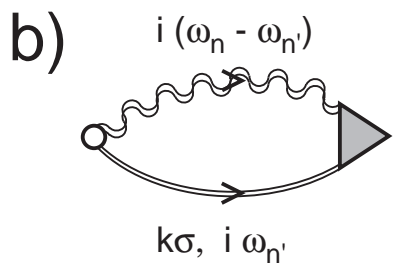
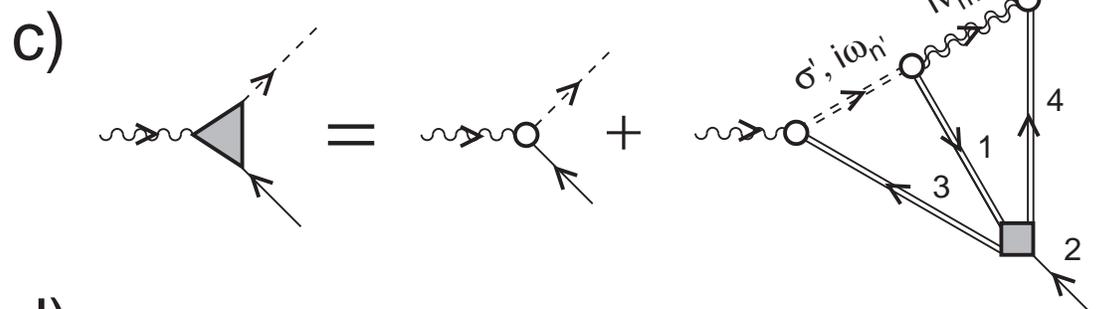
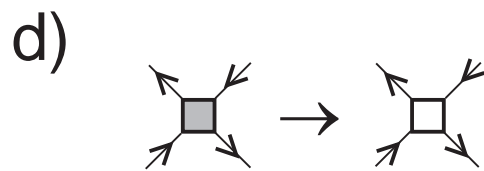
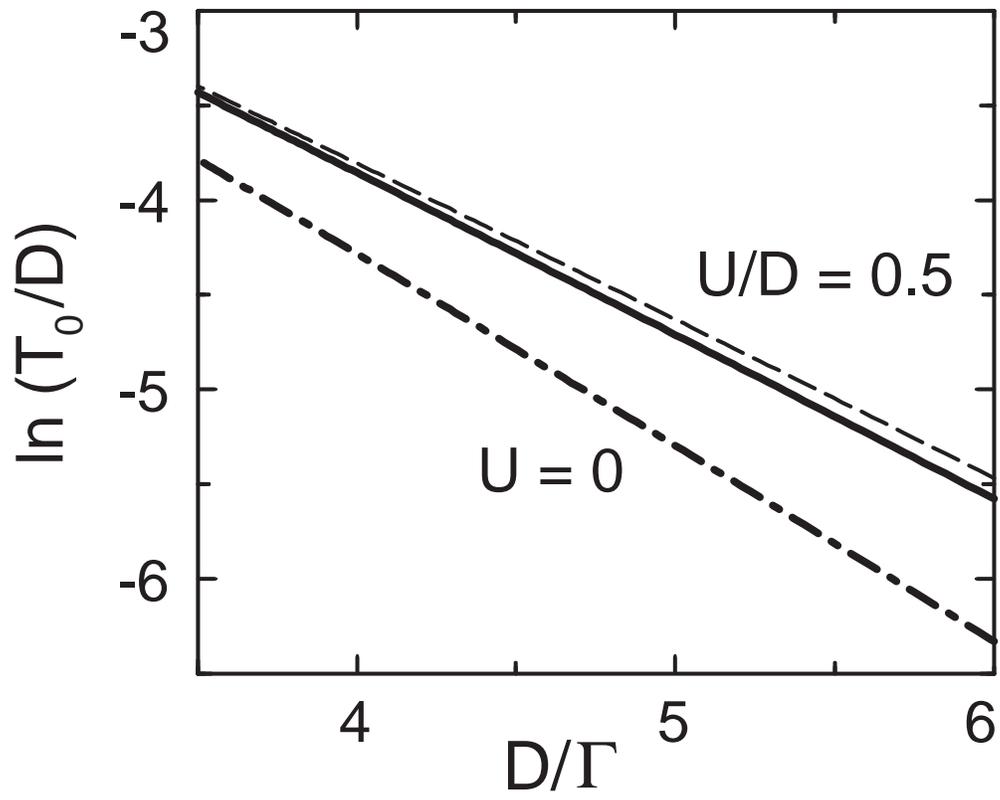
\end{document}